\documentstyle[twocolumn,prl,aps]{revtex}
\tightenlines
\preprint {imsc/98/02/08, cond-mat/9802302}
\begin{document}
\draft
\title{Novel classical ground state of a many body system in arbitrary 
dimensions}

\author{G. Date\footnote{email: shyam@imsc.ernet.in}, Pijush K. 
Ghosh\footnote{email: pijush@imsc.ernet.in}, and M. V. N.
Murthy\footnote{email: murthy@imsc.ernet.in}}

\address
{The Institute of Mathematical Sciences, Madras 600 113, India.\\
}
\date{\today}
\maketitle
\begin{abstract}
The classical ground state of a D- dimensional many body system with two
and three body interactions is studied as a function of the strength of
the three body interaction. We prove {\it exactly} that beyond a critical
strength of the three body interaction, the classical ground state of the
system is one in which all the particles are on a line. The positions of
the particles in this string configuration  are uniquely determined by the 
zeros of the Hermite polynomials. 
\end{abstract}


\narrowtext

In this letter we consider the following Hamiltonian which displays 
remarkable classical properties. The classical Hamiltonian of the model 
for $N$ interacting particles, in $D$ dimensions, is,
\begin{eqnarray}
H &=& \frac{1}{2}\sum_{i=1}^{N}\left[ p_i^{~2} 
+ r_i^2\right] + \frac{g_1}{2}\sum_{i,j (j\ne i)}^N 
\frac{1}{r_{ij}^2} \nonumber \\
&+& \frac{g_2}{2} \sum_{i,j,k (i\ne j\ne k)}^N 
\frac{\vec r_{ij}.\vec r_{ik}}{r_{ij}^2 r_{ik}^2}, \label{ham}
\end{eqnarray}
where $g_1$ and $g_2$ are in general arbitrary but positive coupling 
strengths of the
two and three-body interactions. We have set the mass $m=1$, oscillator 
frequency $\omega=1$ and $\vec r_{ij}=\vec r_i-\vec r_j$. 

We prove exactly that when $g_2 > g_1$ the {\it {classical}} ground
state of this Hamiltonian is unique (modulo rotations) and is described
by all the particles along a line with locations determined by the zeros
of the Hermite polynomial $H_N$.

There are several reasons to consider this Hamiltonian.
The model Hamiltonian given above has a very interesting limit. When 
\begin{equation}
g_1 = g(g+D-2), ~~~~~~ g_2=g^2 ,
\end{equation}
it has been shown that the quantum mechanical ground 
state and an infinite tower of excited
states may be obtained exactly in D-dimensions\cite{khare,marchioro}. ( In 
D=2, some of these  solutions may 
be obtained from the Hamiltonian of the anyons in two dimensions in the
zero total angular momentum sector\cite{date}.) In particular for the
ground state solving the Schroedinger equation $H\psi_0=E_0\psi_0$ we
obtain,
\begin{equation}
\psi_0 = \prod_{i<j} |\vec r_{ij}|^g \exp(-{1\over 2}\sum_i r_i^2),
\label{psigs}
\end{equation}
and the quantum ground state energy is given by,
\begin{equation}
E_0 = {1\over 2}[ND+ g N(N-1)],
\end{equation}
where $g\ge 0$. In the limit $g \rightarrow 0$, the system describes $N$ 
free bosons confined in a harmonic oscillator potential without the 
centrifugal barrier.

For $D = 1$, the model is the celebrated Calogero-Sutherland model
(CSM) \cite{csm}.  The three-body term is identically zero in this 
case. It is known that the classical ground state of CSM is
the one in which the location of the $N$ particles are determined by the 
zeros of the Hermite polynomial $H_N$\cite{hermite}. The CSM is of course 
exactly solvable quantum mechanically.

For $D = 2$, when $g_2=0$, the model reduces to a model with only two-body
inverse square interaction which has some interesting applications to
quantum dots and is considered in detail in ref.\cite{johnson}. Unlike in
one dimension, however, the three body term does not vanish in higher
dimensions and plays a crucial role in the analysis that is to follow.
Similar to the one dimensional CSM the norm of the ground state wave
function in eq.(\ref{psigs}), when $g^2=g_1=g_2$, can be identified with
the joint probability density function of the eigenvalues of complex
random matrices for a particular value of $g$\cite{khare,russia}.  Also, the
ground state wave function is exactly found to be the gauge transformed
Laughlin's wave function and exhibits off-diagonal long range
order\cite{girvin}.  For arbitrary $g_1, g_2$, however, the quantum
problem may not be solvable even for the ground state. 

We also note that the three-body interactions have been used in several
areas like nuclear, atomic and condensed matter physics. The main reason
usually is to account for the many-body effects in systems of particles or
clusters. It is also known that the three-body interactions may be used to
bring about structural transitions in the classical ground state.
Recently, Weber and Stillinger\cite{weber} have proposed a form of the
three-body interaction for inducing structural transitions in the
classical ground state of many particle systems in their molecular
dynamics simulation.  The three-body interaction used here bears a
resemblance to the one proposed by Weber and Stillinger. 

In what follows we keep $g_1$ and $g_2$ arbitrary but positive and analyse
the classical ground state of the Hamiltonian in eq.(\ref{ham}) in 
$D$-dimensions.   
Defining
\begin{equation} 
g_1= g^2, ~~~~ 
g_2=g^2+\lambda, 
\end{equation} 
the Hamiltonian can be written in the form,
\begin{equation}
H = \frac{1}{2}\sum_{i=1}^{N}\left[ p_i^{~2} 
+R_i^2\right] + g\frac{N(N-1)}{2}+\frac{\lambda}{2}V_3, 
\label{ham1}  
\end{equation}
where
\begin{equation}
V_3=\sum_{i,j,k  (i\ne j\ne k)}^N \frac{\vec r_{ij}.\vec r_{ik}}{r_{ij}^2 
r_{ik}^2}, \label{v3} 
\end{equation}
and
\begin{equation}
\vec R_i = \vec r_i - g \sum_{j(j\ne i)} \frac{\vec r_{ij}}{r_{ij}^2}.
\end{equation}
Therefore $\lambda=0$ corresponds to the special case when $g_1=g_2$. 
Here  $g$ 
may be chosen to be either positive or negative. The results that follow,
however, cannot dependent on this choice since the Hamiltonian and the 
equilibrium conditions depend only on $g^2$. We choose to work with $g 
> 0$ since the analysis is simpler. The  
classical equilibrium configurations(fixed points) are obtained by 
solving,      \begin{eqnarray}
\vec p_k &= &0, \\
\sum_{j} \vec R_j.(\nabla_k \vec 
R_j)+\frac{\lambda}{2}\nabla_k V_3 &=&0, \label{eqri}
\end{eqnarray}
where $k=1,...,N$ and the derivatives are taken with respect to 
configuration coordinates $\vec r_k$. 
The energy at any of these extrema are given by, 
\begin{equation} 
E = \sum_{i=1}^{N}\frac{1}{2}R_i^2 + 
g\frac{N(N-1)}{2}+\frac{\lambda}{2}V_3 = \sum_{i=1}^{N} r_i^2, 
\label{eqen}
\end{equation}
where $\vec r_i$ are solutions of the eq.(\ref{eqri}). The last part of 
the above equation follows from the fact that the solutions of the 
equilibrium equations imply, in general,
\begin{equation}
\sum_{i=1}^{N}\frac{1}{2} r_i^2 = \frac{g^2}{2}\sum_{i,j ( j \ne i)} 
\frac{1}{r_{ij}^2} + \frac{1}{2}(g^2+\lambda) \sum_{i\ne j\ne k} 
\frac{\vec r_{ij}.\vec r_{ik}}{r_{ij}^2 r_{ik}^2}.
\end{equation}
Thus the interaction energy is equal to the confinement energy for all 
equilibrium configurations. 

Before analysing the Hamiltonian further we first establish properties of 
the three-body potential $V_3$. Symmetrising $V_3$ explicitly, we have
\begin{equation}
V_3=\frac{1}{3}\sum_{i,j,k  (i\ne j\ne k)}^N 
[\frac{\vec r_{ij}.\vec r_{ik}}{r_{ij}^2 r_{ik}^2}+
\frac{\vec r_{jk}.\vec r_{ji}}{r_{jk}^2 r_{ji}^2}+
\frac{\vec r_{ki}.\vec r_{kj}}{r_{ki}^2 r_{kj}^2}], 
\end{equation}
and using the fact that $\vec r_{ij} +\vec r_{jk} +\vec r_{ki} =0$, it is 
easy to show that 
\begin{eqnarray}
V_3&=&\frac{2}{3}\sum_{i,j,k  (i\ne j\ne k)}^N 
\frac{r_{ij}^2 r_{jk}^2 - (\vec r_{ij}.\vec r_{jk})^2}{r_{ij}^2 r_{jk}^2 
r_{ki}^2} \nonumber \\
&=&\frac{2}{3}\sum_{i,j,k  (i\ne j\ne k)}^N 
\frac{\sin^2\theta_{ijk}}{(\vec r_{ij}+\vec r_{jk})^2}, \label{propv3}
\end{eqnarray}
where $\theta_{ijk}$ is the angle between $\vec r_{ij}$ and $\vec 
r_{jk}$. From eq.(\ref{propv3}) following results follow independent of 
whether $\vec r_i$ are solutions of the equilibrium equations or not: 

\begin{enumerate}
\item Clearly $V_3 \ge 0$ for all $\vec r_i$.

\item Further $V_3=0$  means that each term in 
the summation must be zero since each term is separately a square. 
Therefore, $\theta_{ijk}$ must be either zero or $\pi$. This means that 
all particles must be on  line when $V_3$ attains its global minimum.

\item By a similar calculation it is easy to see that,
$\nabla_i V_3 =0$ for all $i$ if all the particles are on a line, that is,
$\vec r_i = s_i \hat e$, where the unit vector $\hat e$ denotes an 
arbitrary direction in the $D$ dimensional configuration space. 

\item Therefore, if $\vec r_i = s_i \hat e$ is an extremum then the 
solutions of eq.(\ref{eqri}) as well as the energy in eq.(\ref{eqen}) are 
independent of $\lambda$. Furthermore this energy attains its absolute 
minimum value namely,
\begin{equation} 
E_{line}=gN(N-1)/2 ~~~iff~~~\vec R_i =0 ~~~ \forall~~~i. 
\label{eline}
\end{equation}
\end{enumerate}

To compare qualitative features for $\lambda$ positive, negative or zero 
we consider special configurations such that all the particles lie on a 
plane passing through the origin. The plane in fact can be chosen to be 
the x-y plane (say) because of the overall rotational invariance. The 
equilibrium configurations among these are governed by the {\it{same}} 
equations. Of these planer configurations we consider two special ones
which are also solutions of the equilibrium equations, namely,
(1) all N particles on a circle with centre at the origin and (2) (N-1) 
particles on a circle and one at the centre. 
For these configurations the respective energies are given by,
\begin{equation}
E_{\bigcirc}  
= E_{line}\sqrt{1+ \frac{2\lambda}{3g^2}\frac{N-2}{N-1}},
\label{ecirc}
\end{equation}
\begin{equation}
E_{\bigodot}  
= E_{line}\sqrt{1+ 
\frac{2\lambda}{3g^2}\frac{(N+4)(N-3)}{N^2}},
\label{ecdot}
\end{equation}
where $E_{line}$ is given in eq.(\ref{eline}) and refers to the energy of 
the line configuration(which is independent of $\lambda$). 

(a) $\lambda < 0$:  For this case the line configuration {\it {can not}}
attain the absolute minimum of the energy since the special configurations
mentioned above have lower energy. For arbitrary $N$ the ground state  is 
indeed hard to find though it is clear from the form of $V_3$ that these are
necessarily {\it {non one}} dimensional configurations. Among the
{\it {sub-class of planer}} configurations, or equivalently $D=2$, we know 
that 
for $N\le 5$ circle
configuration has the lowest energy where as for $6\le N \le 8$ circle-dot
has the lowest energy. For $N \ge 9 $ multishell configurations have  the
lowest energy \cite{multishell}. For $D > 2$, of course these in general are not
classical ground states.
Note that even though $\lambda <0$,
by definition $\lambda > -g^2$ since the strength of the three body term
is necessarily positive. This ensures that the energy is real in these
configurations. 

(b) $\lambda=0$:  The absolute minimum of energy is attained for all 
configurations satisfying the equation $\vec R_i =0$. The energy in all 
these configurations is given by,
\begin{equation}
E=g\frac{N(N-1)}{2} = E_{line}.
\end{equation}
In this case the line, the circle and the circle-dot are all degenerate. 
Within the sub-class of planer configuration (equivalently for $D = 2$) 
numerical simulations show that there are many other configurations which 
are also degenerate. For $D = 2$, the quantum mechanical ground state can 
be obtained exactly in this limit as noted at the beginning. The classical 
energy is the same as the quantum mechanical ground state energy apart 
from the contribution from the zero point fluctuation. Note that this is 
rather special for $D=2$. 

(c)$\lambda >0$: The line configuration is certainly a ground state since 
the absolute minimum of the energy is attained. Because $V_3$ is positive 
definite or zero, any other extremum must necessarily have higher energy, 
\begin{equation}
E \ge g \frac{N(N-1)}{2}=E_{line}~~~\forall~~~\lambda \ge 0 .
\end{equation}

Thus for all $\lambda >0$, the classical ground states are line 
configurations with $\vec r_i = s_i \hat e$, where $\hat e$ defines an 
arbitrary direction in the $D$ dimensional space. The positions of the 
particles on the line, $s_i$ are determined by the equation $\vec R_i=0$. 
That is,
\begin{equation}
s_i = g \sum_{j(j\ne i)}^N \frac{1}{s_{ij}} ~~~\forall ~~~ i=1,...,N  
\label{eqsi} 
\end{equation}
and an absolute minimum is attained when each $s_i$ satisfies this equation. 

We now show that eqs.(\ref{eqsi}) in fact have a unique 
solution. We scale out $g$ by defining a new variable, $t_i=g^{-1/2} s_i$. 
Define 
\begin{equation}
C_k = \sum_{i=1}^N t_i^{k+1}.
\end{equation}
Using the equilibrium condition,
\begin{equation}
C_k = \frac{1}{2} \sum_{i,j(i\ne j)} \frac{t_i^k -t_j^k}{t_{ij}}
= \frac{1}{2} \sum_{i,j(i\ne j)}\sum_{r=0}^{k-1} t_i^r t_j^{k-1-r}. 
\end{equation}
Manipulating the summations, we get the recursion relation,
\begin{equation}
C_{k+1} =\frac{1}{2}[ \sum_{r=0}^{k} C_{r-1} C_{k-1-r} -(k+1) 
C_{k-1}]~~~\forall ~~~k\ge 0. 
\end{equation}
$C_k$ are clearly function of only $N$ and are independent of any 
particular solution of eq.(\ref{eqsi}). Also, by definition $C_{-1}=N$ 
and $C_0=0$ using directly the eq.(\ref{eqsi}). 

Let $t'_i$ be another solution of the eq.(\ref{eqsi}). Because the $C_k$ 
are independent of the particular solution, we have,
\begin{equation}
C_k = \sum_{i=1}^N t_i^{k+1}=\sum_{i=1}^N (t')_i^{k+1}.
\end{equation}
Therefore,
\begin{equation}
\sum_{i=1}^N (t_i^{k+1}-(t')_i^{k+1}) = \sum_i \Delta t_i a_{ki}=0,
\end{equation}
where $\Delta t_i = t_i-(t')_i$ and $a_{ki}=\sum_{r=0}^{k} t_i^r 
(t')_i^{k-r}$. Multiplying both sides by arbitrary $b_k$ and summing over k 
we get,
\begin{equation}
\sum_i \Delta t_i \beta_i=0,
\end{equation}
where $\beta_i = \sum_{k=0}^{\infty} a_{ki}b_k$. Since $b_k$ and 
therefore $\beta_i$ are completely arbitrary we have $\Delta t_i=0$. 
Therefore the solution is unique. It is well known\cite{hermite}, 
in fact, the 
N-zeros of the Hermite polynomial $H_N(t)$, namely $t_i$, satisfy the 
eq.(\ref{eqsi}) 
with $t_i = g^{-1/2}s_i$. Hence the uniqueness proof implies that the 
positions of the particles along a line in the ground state are 
determined by the zeros of the Hermite polynomial.  

As an aside, we also remark that since $C_k$ are constructed 
out of the zeros of the Hermite polynomials, we have also obtained an 
infinite set of identities for the zeros of the Hermite polynomial. 

This completes the proof that for all $\lambda >0$ there is only one
unique global minimum (modulo the states obtained by rotations). The
corresponding configuration is a line where the positions of the particles
are given as the zeros of Hermite polynomial in terms of the scaled
variables $t_i$. 

We have thus demonstrated that a $D$ dimensional Hamiltonian with two and
three body interactions has a unique stringlike ground state when the
strength of the three body term exceeds a critical value determined in
terms of the parameter $\lambda$. Further we note:
\begin{enumerate}

\item The result is independent of the strength of the two-body term, the
number of particles and the strength of the oscillator potential. Thus
even in the full quantum theory it is possible to choose $g$ or $N$ large
enough such that the quantum fluctuations about the classical ground state
are as small as we wish. The above statements about the classical ground
state approximately remain valid, therefore, even in the quantum theory. 

\item The string like ground state one obtains here for $\lambda > 0$ is
identical to the classical ground state of the one-dimensional CSM
Hamiltonian. 

\item The model provides an example of a structural transition driven by
the mutual interactions without any external influence. In fact this
statement is manifestly independent of the space dimension.
Thus we have a powerful result that in any space-dimension, the
Hamiltonian in eq.(1) gives rise to string like ground state when the
strength of the three body term exceeds the strength of the two body term
irrespective of the number of particles, the strength of
the confinement potential and the strength of the two body term.

\item Finally, we point out that the form of the three-body potential as
given in eq.(\ref{propv3}), may be easily generalised. Note that the
results of the analysis remain unchanged as long as $V_3$ and its gradient
vanish on a line and $V_3$ is positive semi-definite. In particular, 
\begin{eqnarray}
v_3 (\vec{r}_{ij}, \vec{r}_{jk}, \vec{r}_{ki}) & = & 
(\vec{r}_{ij} \cdot \vec{r}_{jk})(\vec{r}_{jk} \cdot \vec{r}_{ki}) 
 + cyclic \nonumber \\
& = & r_{ij}^2 r_{jk}^2 \sin^2(\theta_{ijk})
\end{eqnarray}
satisfies these properties. 
Hence it follows that for any $V_3$ of the form, 
\begin{equation} 
V_3 = \sum_{i\ne j\ne k} f(r_{ij}^2,r_{jk}^2,r_{ki}^2)
v_3 (\vec{r}_{ij}, \vec{r}_{jk}, \vec{r}_{ki})  
\end{equation} 
where $f(r_{ij},r_{ik},r_{jk}) $ is a symmetric function of its arguments
and is positive,  also vanishes for a line together with its gradient. 
The $V_3$ considered in eq.(\ref{propv3}) corresponds to 
$$
f(r_{ij}, r_{jk}, r_{ki}) ~ = ~ (r_{ij}^2 r_{jk}^2 r^2_{ki})^{-1} 
$$ 
Thus, 
for any arbitrary $f$ modulo these properties, the mere presence of a 
three-body interaction is sufficient to give a string like ground state. 
The crucial point here is that the geometric structure of the ground 
state depends on the $\sin^2(\theta_{ijk})$ which favours string 
configurations and not on the precise form of $f$.
The three-body interaction here differs from the one proposed 
by Weber and Stillinger\cite{weber}, in that $2\theta_{ijk}$ appears 
in place of $\theta_{ijk}$. While the final ground state configuration 
may be different in these two cases, the presence of the three body term 
is crucial in defining the actual ground state configuration.

\end{enumerate}

\acknowledgments 
We thank A. Khare for suggestions and a critical reading of the
manuscript. We also thank G. Baskaran, R. K. Bhaduri, P. P. Divakaran,
Tabish Qureishi, Madan Rao, R. Shankar and Surajit Sengupta for 
discussions and comments. 


\end{document}